\documentclass[aps,prl,twocolumn,superscriptaddress,amsmath,nofootinbib,natbib]{revtex4-1}

\usepackage[hidelinks]{hyperref}
\usepackage{amsfonts}
\usepackage{mathtools}
\usepackage{aas_macros}
\usepackage[capitalize]{cleveref}                                                             
\usepackage[export]{adjustbox}
\usepackage{lipsum}

\Crefname{equation}{Equation}{Equations}
\crefname{equation}{Eq.}{Eqs.}
\Crefname{figure}{Figure}{Figures}
\crefname{figure}{Fig.}{Figs.}
\Crefname{table}{Table}{Tables}
\crefname{table}{Tab.}{Tabs.}
\Crefname{section}{Section}{Sections}
\crefname{section}{Sec.}{Secs.}

\renewcommand{\d}[2][]{\operatorname{d}^{#1}\!{#2}}

\begin{document}

\title{Curvature tension: evidence for a closed universe}

\author{Will Handley}
\email[]{wh260@mrao.cam.ac.uk}
\affiliation{Astrophysics Group, Cavendish Laboratory, J.J.Thomson Avenue, Cambridge, CB3 0HE, UK}
\affiliation{Kavli Institute for Cosmology, Madingley Road, Cambridge, CB3 0HA, UK}
\affiliation{Gonville \& Caius College, Trinity Street, Cambridge, CB2 1TA, UK}
\homepage[]{https://www.kicc.cam.ac.uk/directory/wh260}

\date{\today}

\begin{abstract}
    The curvature parameter tension between Planck 2018, cosmic microwave background lensing, and baryon acoustic oscillation data is measured using the suspiciousness statistic to be $2.5$ to $3\sigma$. Conclusions regarding the spatial curvature of the universe which stem from the combination of these data should therefore be viewed with suspicion. Without CMB lensing or BAO, Planck 2018 has a moderate preference for closed universes, with Bayesian betting odds of over $50:1$ against a flat universe, and over $2000:1$ against an open universe.
\end{abstract}

\maketitle
\section{Introduction}\label{sec:introduction}

Quantifying the consistency between different cosmological datasets has become increasingly important in recent years. Observations of the early and late time universe give undeniably inconsistent predictions for the present day value of the expansion rate~\cite{HubbleTension}, a phenomenon known as the ``Hubble tension''. This paper explores a second ``curvature tension'' present within early-time observations.

Despite the long history of cosmological models which include spatial curvature~\cite{FLRW1,FLRW2,FRLW3,FRLW4}, in the modern era there is a strong research community bias toward a flat universe. This is partly on theoretical grounds, since the prevailing (and very successful) inflationary theory of the primordial  universe predicts a late-time cosmos which is extremely close to flat~\cite{starobinskii_spectrum_1979,guth_inflationary_1981,linde_1982}.  Curved models are also likely under-represented in the literature due in large part to the increased theoretical and numerical computational cost associated with curved cosmologies\footnote{The results presented in this paper required over twelve years of high-performance computing time (or two weeks on 320 cores)}.

\begin{figure*}
    \includegraphics{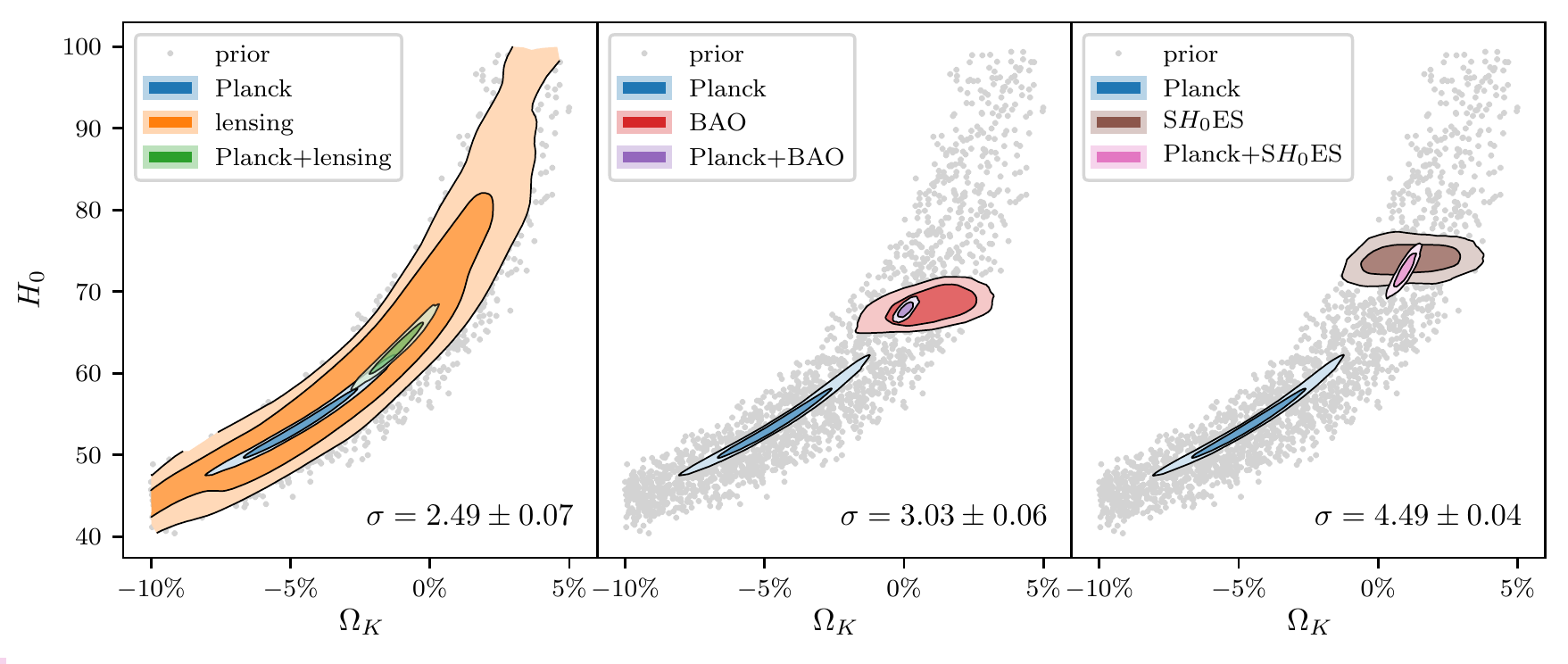}
    \caption{Three tensions from \cref{fig:tensions} plotted in the $(\Omega_K,H_0)$ plane using \texttt{anesthetic}~\cite{anesthetic}. In the first panel, whilst the lensing posterior alone is barely distinguishable from the prior, when combined with Planck, lensing has draws the combined posterior significantly toward flatness, at a tension of 2.5$\sigma$. The second panel shows BAO's preference for a flat universe. The BAO posterior is disconnected from the Planck posterior, at a tension of $3\sigma$. Finally, in the third panel the Planck-S$H_0$ES inconsistency in the curved case is shown to be 4.5$\sigma$.\label{fig:omegak_H0}}
\end{figure*}

However, most of the bias toward flat cosmologies is undoubtedly derived from observational considerations, which ever since COBE~\cite{COBE} and BOOMERANG~\cite{BOOMERANG} have confirmed that the universe is consistent with a flat cosmology to within $\Omega_K=\pm 10\%$. The primary conclusion of this paper is that such a position can no longer be consistently held, as Planck 2018 cosmic microwave background data~\cite{parameters,likelihoods} alone suggests a model that is closed at $\Omega_K\sim -4.5\% \pm 1.5\%$, with betting odds of $50:1$ against a flat universe. Other datasets such as Planck 2018 CMB lensing~\cite{lensing} or baryon acoustic oscillations~\cite{SDSS,SDSS2,SDSS3} which strongly suggest a flat universe are quantitatively inconsistent at 2.5 to 3$\sigma$ with CMB data alone, and should not be confidently combined until this tension is released.
Results are visualised in \cref{fig:omegak_H0} and summarised in \cref{fig:tensions,fig:evidences}.

\section{Background}\label{sec:background}

\subsection{Cosmology with curvature ($K\Lambda$CDM)}

Under the extended Copernican principle, the universe is assumed at zeroth order to be homogeneous and isotropic at the largest scales. Einstein's theory of general relativity under these assumptions yields the Friedmann--Lema{\^i}tre--Robertson--Walker cosmology~\cite{FLRW1,FLRW2,FRLW3,FRLW4}. The metric in reduced spherical polar coordinates takes the form
\begin{equation}
    \hspace{-4pt}\d{s}^2 = \d{t}^2 - a(t)^2 \left[ \frac{\d{r}^2}{1-Kr^2} + r^2(\d{\theta}^2 + \sin^2\theta \d{\phi}^2) \right],
    \label{eqn:metric}
\end{equation}
whereby homogeneous and isotropic spatial slices expand or contract over cosmic time $t$ via the scale factor $a$. The spatial slices come in one of three forms, flat Euclidean space ($K=0$), open hyperbolic space ($K=-1$) or closed hyperspherical space ($K=+1$). Curvature is typically quantified via its fractional contribution to the cosmic energy budget today $\Omega_K$, which is measured as a percentage and has the opposite sign to $K$.
In this work, I follow the notation of~\cite{powerspectrum} and denote the concordance cosmology with $K=0$ as $\Lambda$CDM, and it's extension with $\Omega_K$ considered as an unknown parameter as $K\Lambda$CDM.

\subsection{Bayesian statistics}

Given a set of data $D$ and a predictive model $M$ with parameters $\theta$,
Bayes' theorem relates the statistical inputs to inference (the likelihood and prior) to the statistical outputs (the posterior and evidence):
\begin{align}
    P(D|\theta,M) \times P(\theta|M) &= P(\theta|D,M) \times P(D|M), \nonumber\\
    \text{Likelihood} \times \text{Prior} &= \text{Posterior} \times \text{Evidence},
    \label{eqn:bayes_theorem}
 \\
    \mathcal{L} \times \pi &= \mathcal{P} \times \mathcal{Z}.\nonumber
\end{align}
The posterior is the central quantity in parameter estimation (what the data tell us about parameters of a model), whilst the evidence is pivotal in model comparison (what the data tell us about the relative quality of a model)~\cite{Skilling2006,Trotta2008}. The evidence ratio gives the Bayesian betting odds one would assign between two competing models, assuming equal {\em a priori\/} model probability.

Parameter estimation is typically performed by compressing the high-dimensional posterior into a set of representative samples via a Markov-Chain Monte-Carlo process~\cite{cosmomc}. The evidence is derived from the likelihood by marginalising over the prior
\begin{equation}
    \mathcal{Z}=\int \mathcal{L}(\theta)\pi(\theta)\d{\theta} = \left\langle \mathcal{L}\right\rangle_\pi,
    \label{eqn:evidence}
\end{equation}
which may be computed numerically via a Laplace approximation~\cite{mackay2003information}, Savage Dickey ratio~\cite{SavageDickey}, nearest-neighbour volume estimation~\cite{MCEvidence1,MCEvidence2} or nested sampling~\cite{Skilling2006,Feroz2008,PolyChord0,PolyChord1,DNest,DNest4,dynesty}. If one has access to the evidence, then it is straightforward to compute the Kullback-Leibler divergence
\begin{equation}
    \mathcal{D} = \int \mathcal{P}(\theta) \log \frac{\mathcal{P}(\theta)}{\pi(\theta)}\d{\theta} = \left\langle \log\frac{\mathcal{P}}{\pi}\right\rangle_\mathcal{P},
    \label{eqn:KL}
\end{equation}
which quantifies the degree of compression from prior to posterior provided by the data.

As a model comparison tool, the evidence naturally quantifies Occam's razor, incorporating a parameter volume-based penalty that penalises models with unnecessary constrained parameters. This penalty factor may be approximated using the difference in KL divergence between models. For further detail on Bayesian statistics, readers are recommended references~\cite{hobson2010bayesian,Trotta,mackay2003information,sivia2006data}.

\begin{figure*}
    \includegraphics{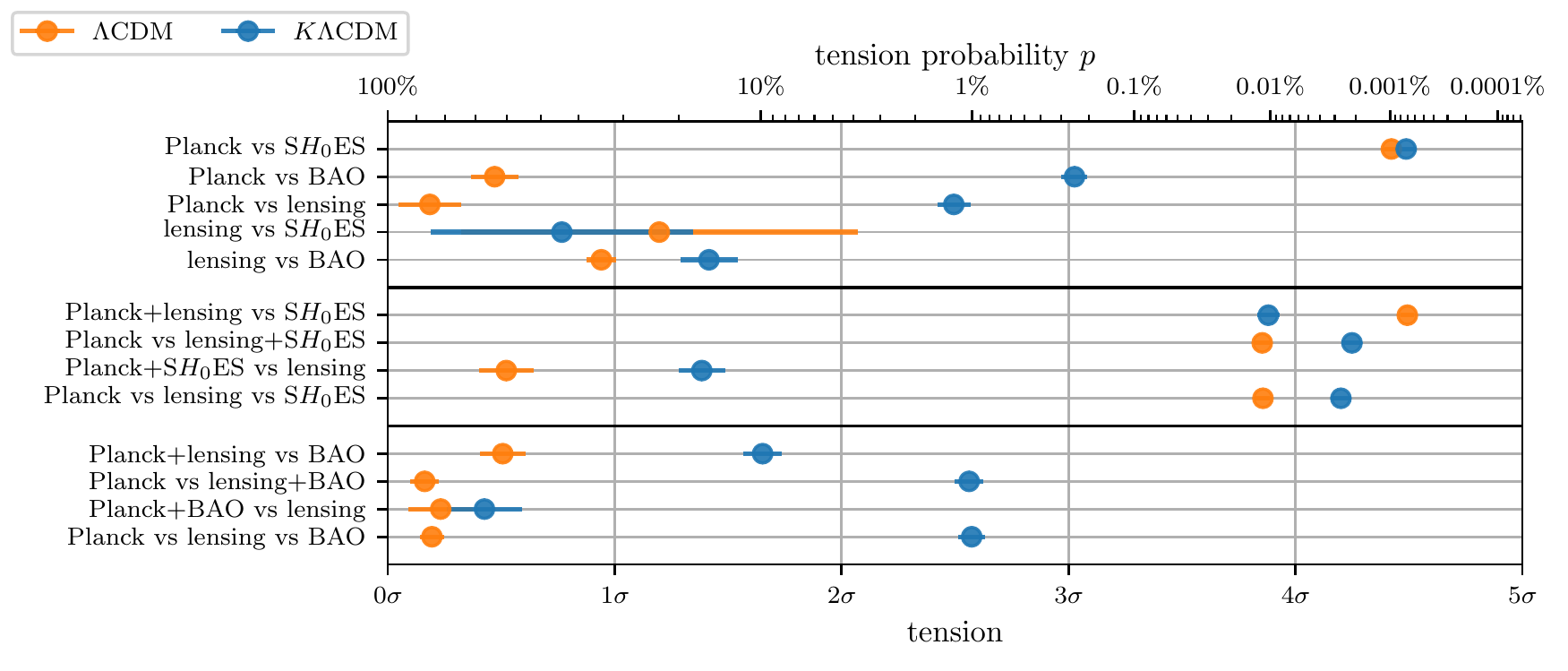}
    \caption{Parameter tensions computed using \texttt{anesthetic}~\cite{anesthetic}. The top line shows the severe tension between S$H_0$ES and Planck detailed in~\cite{tension,dimensionality} updated for Planck 2018 data. For $\Lambda$CDM Planck, lensing and BAO are all consistent. For $K\Lambda$CDM, Planck and BAO are strongly inconsistent, whilst Planck and lensing are moderately inconsistent. The ninth and final rows indicate the triple tension from \cref{eqn:triple_tension}, and show strong mutual inconsistency between Planck, lensing and S$H_0$ES, and moderate mutual inconsistency between Planck, lensing and BAO. Curvature in general enhances tension, but relaxes it for Planck+lensing vs S$H_0$ES. The large error bars for lensing vs S$H_0$ES occur due to the fact their shared constrained parameters have $\tilde{d}\ll1$. \label{fig:tensions}}
    \vspace{-10pt}
\end{figure*}

\subsection{Tension quantification}
The Bayesian evidence from \cref{eqn:evidence} also appears in tension quantification~\cite{tension,Marshall2006,Verde2013,Verde2014,Raveri2016a,Seerhars2016,DESParameters2017}. The Bayes ratio
\begin{align}
    R = \frac{P(D_2|D_1)}{P(D_2)} = \frac{P(D_1|D_2)}{P(D_1)} = \frac{\mathcal{Z}_{12}}{\mathcal{Z}_1 \mathcal{Z}_2},
    \label{eqn:R}
\end{align}
quantifies the compatibility of two datasets $D_1$ and $D_2$ by giving a Bayesian's relative confidence in their combination.
As with many Bayesian quantities, $R$ is naturally prior dependent~\cite{tension}. In the absence of well-motivated priors, or if deliberately over-wide ranges on parameters are used (as is often the case in cosmology), the prior dependency can be removed from $R$ using the KL divergence from \cref{eqn:KL}. Dividing $R$ by the information ratio $I$ gives the suspiciousness $S$
\begin{equation}
    S = \frac{R}{I}, \qquad \log I = \mathcal{D}_1 + \mathcal{D}_2 - \mathcal{D}_{12},
    \label{eqn:S}
\end{equation}
Using a Gaussian analogy, one may calibrate $S$ into a tension probability $p$ and convert this into an equivalent ``sigma value'' via the survival function of the chi-squared distribution
\begin{equation}
    p = \int_{d-2\log S}^\infty \chi_d^2(x)\d{x}, \qquad \sigma= \sqrt{2} \mathrm{Erfc}^{-1}(p),
\end{equation}
where $\mathrm{Erfc}^{-1}$ is the inverse complementary error function and $d$ is quantified by the Bayesian model dimensionality~\cite{dimensionality,gelman,Spiegelhalter2,Kunz,information_criteria}
\begin{equation}
    d =\tilde{d}_1 +\tilde{d}_2- \tilde{d}_{12}, \qquad\tilde{d}/2 = \langle {\left(\log\mathcal{L}\right)} ^2 \rangle_\mathcal{P} - \langle \log\mathcal{L} \rangle_\mathcal{P}^2.
    \label{eqn:d}
\end{equation}
The differences in model dimensionalities recover the number of shared constrained parameters.
See~\cite{tension,dimensionality} for a more detailed discussion of tension and model dimensionality. Other tension metrics are also available~\cite{Charnock:2017,Inman1989,Battye2015,Seehars2014,Nicola2019,Kunk2006,Karpenka2015,MacCrann2015,Adhikari2019,Douspis,Raveri2018,Hobson2002}.

\begin{table}
    \begin{tabular}{lp{180pt}}
        \hline
        Abbrv. & Description \\
        \hline
        \hline
        Planck~\cite{likelihoods} &
        Temperature and polarization anisotropies in the CMB measured from the publicly available baseline Planck 2018 TTTEEE+lowE likelihood. \\
        lensing~\cite{lensing} &
        Planck 2018 baseline CMB lensing likelihood. \\
        BAO~\cite{SDSS,SDSS2,SDSS3} & Baryon Acoustic Oscillation and Redshift Space Distortion measurements from the Baryon Oscillation Spectroscopic Survey (BOSS) DR12. \\
        S$H_0$ES~\cite{Riess2018}&
        Supernovae and $H_0$ for the Equation of State:
        local cosmic distance ladder measurements of the expansion rate, using type Ia SNe calibrated by variable Cepheid stars. Implemented as a Gaussian likelihood.
        \\
    \end{tabular}
    \caption{Abbreviations, citations and descriptions for observational datasets used in this analysis.\label{tab:datasets}}
\end{table}

In this work the ``triple tension'' is also introduced to quantify the tension between three datasets simultaneously, extending the definition of $R$, $I$ and $d$ from \cref{eqn:R,eqn:S,eqn:d} when necessary to
\begin{equation}
    R = \frac{\mathcal{Z}_{123}}{\mathcal{Z}_1\mathcal{Z}_2\mathcal{Z}_3}, \quad  
    \begin{array}{c}
        \log I = \mathcal{D}_1 + \mathcal{D}_2 + \mathcal{D}_3- \mathcal{D}_{123}, \\
        d =\tilde{d}_1 +\tilde{d}_2+\tilde{d}_3- \tilde{d}_{123},
    \end{array}
    \label{eqn:triple_tension}
\end{equation}
with obvious generalisation to four or more parameters.

\section{Methodology}\label{sec:methodology}

The observational datasets used for this analysis are described in \cref{tab:datasets}. Note that throughout this paper ``lensing'' is an abbreviation for ``Planck CMB lensing''.

\begin{figure}
    \includegraphics[right]{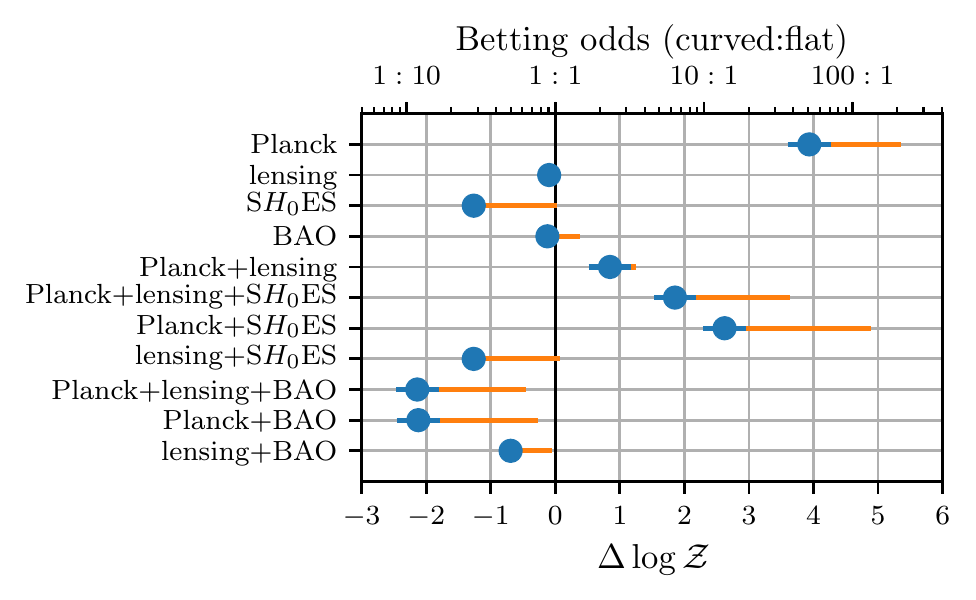}
    \caption{Model comparison between curved $K\Lambda$CDM and flat $\Lambda$CDM cosmologies. Positive $\Delta\log\mathcal{Z}$ indicates favouring of curved universes. The Occam regularisation penalty associated with the additional $\Omega_K$ parameter is shown in orange, estimated via the difference in KL divergence between the two models. \label{fig:evidences}}
\end{figure}

Bayesian evidences and posteriors are computed using nested sampling~\cite{Skilling2006} provided by \texttt{CosmoChord}\cite{CosmoChord}, an extension of \texttt{CosmoMC}~\cite{cosmomc} using \texttt{PolyChord}~\cite{PolyChord0,PolyChord1} to sample efficiently in high dimensions and exploit the fast-slow cosmological parameter hierarchy~\cite{cosmomc_fs}. The post-processing computations for the evidence, KL divergence and Bayesian model dimensionality detailed in \cref{fig:tensions,fig:evidences}, as well as the posterior plots in \cref{fig:omegak_H0,fig:constraints} are computed using the \texttt{anesthetic}~\cite{anesthetic} software package.

All cosmological and nuisance parameters are varied in the nested sampling runs. The cosmological prior widths (visualised in \cref{fig:constraints}) are narrower than the \texttt{CosmoMC} defaults. This is necessary since nested sampling begins by exploring the deep tails of the distribution. In these regions unphysical and unforeseen combinations of parameters causes cosmology codes to fail. Whilst flat cosmology codes have been extensively stress-tested by nested sampling, the curved branches have not. Since the priors are wide enough to fully encompass the posteriors, there would be little quantitative effect on the conclusions of this paper from using a wider set of priors.

All inference products required to compute results are available for download from Zenodo~\cite{Zenodo}.

\section{Results}\label{sec:results}
Tension results are shown in \cref{fig:tensions}. For $\Lambda$CDM, Planck, lensing and BAO are all consistent, with tensions $\le2\sigma$. For $K\Lambda$CDM, whilst BAO and lensing are consistent with one another, they are inconsistent with Planck at $\ge2.5\sigma$, a tension similar to that between Planck 2015 and weak galaxy lensing (DES)~\cite{tension,DESParameters2017}. The Hubble tension between S$H_0$ES and Planck measured using the suspiciousness statistic is found to be $\sim4.5\sigma$. Adding curvature generally enhances all tensions considered, except for Planck+lensing vs S$H_0$ES, for which the tension is weakened due to the increased error bar on $H_0$ from the inclusion of curvature.

Model comparison results are shown in \cref{fig:evidences}. Without lensing, Planck prefers curved universes at a ratio of $50:1$ with $\Omega_K = -4.5\% \pm 1.5\% $. Examining the posterior in \cref{fig:omegak_H0}, only $1/2000$ of the posterior mass has $\Omega_K>0$ indicating a strong preference for closed universes over open ones. When the moderately inconsistent lensing data are added, $K\Lambda$CDM is only very slightly preferred to the concordance model at $2:1$, but still prefers closed universes with $\Omega_K =-1.2\%\pm0.6\% $. If one discounts the strong discordance of BAO and includes it in model constraints, flat universes become preferred in a Bayesian sense due to the effect of the Occam penalty penalising the additional constrained parameter consistent with $\Omega_K=0.0\%\pm0.2\%$, degenerate with a flat universe. 

\begin{figure*}
    \includegraphics{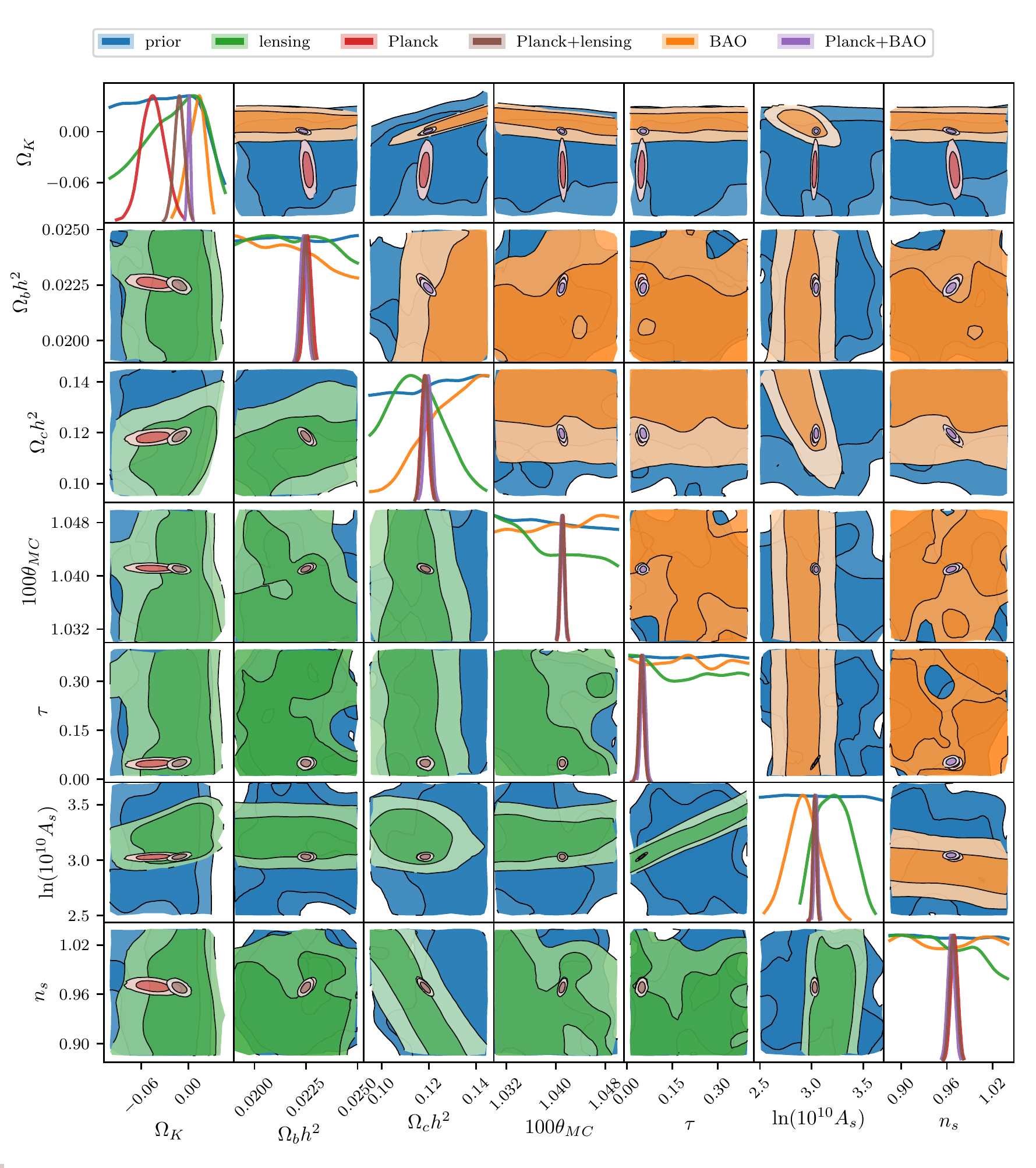}
    \caption{One- and two-dimensional marginalised priors and posteriors for the seven cosmological parameters of the $K\Lambda$CDM cosmology, plotted with \texttt{anesthetic}~\cite{anesthetic}. Plots on the diagonal are one-dimensional marginalisations, and off-diagonal plots are contour plots showing marginal iso-probability contours containing 68\% and 95\% of the marginalised posterior mass. The prior in blue and the Planck posterior in red are included in all plots, below-diagonal elements show lensing posteriors, above-diagonal elements show posteriors for BAO. The left-most column and top-most row detail curvature, where the tension between the posteriors can be seen clearly. The remaining parameters show no discernible tensions. Lensing constrains approximately two parameters, a linear combination of $\Omega_ch^2$ and $n_s$, and a linear combination of $\ln(10^{10}A_s)$ and $\tau$, which is reflected in the Bayesian model dimensionality of $\tilde{d}_\mathrm{lensing}=1.8\pm1$. BAO has a model dimensionality of $\tilde{d}_\mathrm{BAO}=2.6\pm0.1$, consistent with its constraints on $\ln(10^{10}A_s)$ and $\Omega_K$, and its partial constraint on $\Omega_ch^2$.\label{fig:constraints}}
\end{figure*}

The marginal distribution for the Planck, lensing and BAO likelihoods for all cosmological parameters are shown in \cref{fig:constraints}.

\section{Conclusions}\label{sec:conclusions}

In light of the inconsistency between Planck, CMB lensing and BAO data in the context of curved universes, cosmologists can no longer conclude that observations support a flat universe. If one assumes Planck CMB data are correct, then we should conclude with odds of $50:1$ that the universe is closed.

These results by no means prove that the universe is curved. As with the Hubble tension, it may be that there are unaccounted systematic errors, bugs in some or all of the curved likelihood codes, or the possibility of new physics that could eventually release the parameter tension in $\Omega_K$. What one can unambiguously say is that further research is required to uncover why the CMB alone so strongly prefers a closed universe, whilst other datasets provide quantitatively contradictory constraints.

The observational probes and codes used in this work are all nominally curvature-agnostic. However, since cosmological software packages and data compression mechanisms often require approximations, fiducial biases toward flatness can easily creep in. A thorough investigation of these implicit assumptions forms the natural follow-up to this work. 

Closed universe models can generally relax the Hubble tension between supernovae observations and the CMB. If the solution to curvature tension resides in new physics, then it is possible that an additional change alongside the introduction of curvature may yet resolve both the Hubble and the curvature inconsistencies. The fact that curved cosmologies relax the strong CMB constraint on the Hubble constant may aid additional modifications in finding a resolution to the Hubble tension.

\section{Acknowledgements}
\begin{acknowledgements}
This work was performed using resources provided by the \href{http://www.csd3.cam.ac.uk/}{Cambridge Service for Data Driven Discovery (CSD3)} operated by the University of Cambridge Research Computing Service, provided by Dell EMC and Intel using Tier-2 funding from the Engineering and Physical Sciences Research Council (capital grant EP/P020259/1), and \href{www.dirac.ac.uk}{DiRAC funding from the Science and Technology Facilities Council}.

I thank Gonville \& Caius College for their support via a Research Fellowship. 
I am also grateful to Lukas Hergt for invaluable contributions to the \texttt{anesthetic} package. I thank Pablo Lemos, Will Coulton, Oliver Friedrich, Liam Lau, Anthony Lasenby, Mike Hobson, Hiranya Peiris and Benjamin Joachimi for useful conversations regarding tension and curvature.
\end{acknowledgements}

\bibliographystyle{unsrtnat}
\bibliography{curvature_tension.bib}

\end{document}